\documentclass[sigconf,natbib=true,screen=true]{acmart}

\copyrightyear{2026}
\acmYear{2026}
\setcopyright{cc}
\setcctype{by}
\acmConference[SIGIR '26]{Proceedings of the 49th International ACM SIGIR Conference on Research and Development in Information Retrieval}{July 20--24, 2026}{Melbourne, VIC, Australia}
\acmBooktitle{Proceedings of the 49th International ACM SIGIR Conference on Research and Development in Information Retrieval (SIGIR '26), July 20--24, 2026, Melbourne, VIC, Australia}
\acmDOI{10.1145/3805712.3809746}
\acmISBN{979-8-4007-2599-9/2026/07}

\usepackage{graphicx}
\usepackage{booktabs}
\usepackage{subcaption}
\usepackage{multirow}
\usepackage{float}
\usepackage{amsmath, amsfonts}
\usepackage{dsfont}
\usepackage{threeparttable}
\usepackage{tabularx}
\usepackage{xspace}
\usepackage[inline]{enumitem}
\usepackage{framed}
\usepackage{booktabs}

\usepackage[table]{xcolor}
\usepackage{adjustbox}
\usepackage{placeins}

\usepackage[ruled,vlined,linesnumbered]{algorithm2e}
\DontPrintSemicolon
\SetKwInput{KwIn}{Input}
\SetKwInput{KwOut}{Output}
\SetKw{KwRet}{return}
\usepackage{etoolbox}

\begin{document}

\title{Inductive Dual-Polarity Modeling via Static–Dynamic Disentanglement for Dynamic Signed Networks}

\author{Yikang Hou}
\orcid{0009-0009-7696-140X}
\affiliation{%
  \institution{Southwest University}
  \department{College of Computer and Information Science}
  \city{Chongqing}
  \country{China}}
\email{jh0321@email.swu.edu.cn}

\author{Junjie Huang}
\authornote{Corresponding author.}
\orcid{0009-0002-2046-5393}
\affiliation{%
  \institution{Southwest University}
  \department{College of Computer and Information Science}
  \city{Chongqing}
  \country{China}}
\email{junjiehuang@swu.edu.cn}

\author{Yijun Ran}
\orcid{0000-0002-7047-3343}
\affiliation{%
  \institution{Guizhou Normal University}
  \department{School of Big Data and Computer Science}
  \city{Guiyang}
  \country{China}}
\email{yijunran@gznu.edu.cn}

\author{Tao Jia}
\orcid{0000-0002-2337-2857}
\affiliation{%
  \institution{Southwest University}
  \department{College of Computer and Information Science}
  \city{Chongqing}
  \country{China}
}
\affiliation{
  \institution{Chongqing Normal University}
  \department{College of Computer and Information Science}
  \city{Chongqing}
  \country{China}
}
\email{tjia@swu.edu.cn}

\begin{abstract}
Dynamic signed networks (DSNs) are common in online platforms, where time-stamped positive and negative relations evolve over time. A core task in DSNs is dynamic edge prediction, which forecasts future relations by jointly modeling edge existence and polarity (positive, negative, or non-existent). However, existing dynamic signed network embedding (DSNE) methods often entangle positive and negative signals within a shared temporal state and rely on node-specific temporal trajectories, which can obscure polarity-asymmetric dynamics and harm inductive generalization, especially under cold-start evaluation.
We study an inductive setting where each test edge contains at least one endpoint node held out from training, while its interactions prior to the prediction time are available as historical evidence. The model must therefore infer representations for unseen nodes solely from such limited history. We propose \textbf{IDP-DSN}, an \textbf{I}nductive \textbf{D}ual-\textbf{P}olarity framework for \textbf{D}ynamic \textbf{S}igned \textbf{N}etworks. IDP-DSN maintains sign-selective memories to model positive and negative temporal dynamics separately, performs history-only neighborhood inference for unseen nodes (instead of learned node-wise trajectories), and enforces polarity-wise static--dynamic disentanglement via an orthogonality regularizer.
Experiments on BitcoinAlpha, BitcoinOTC, Wiki-RfA, and Epinions demonstrate consistent improvements over the strongest baselines, achieving relative Macro-F1 gains of \textbf{16.8/23.4\%}, \textbf{16.9/24\%}, \textbf{30.1/25.5\%}, and \textbf{18.7/28.9\%} in the transductive/inductive settings, respectively. These results highlight the effectiveness of IDP-DSN on DSNs, particularly under inductive cold-start evaluation for dynamic signed edge prediction.\footnote{Our code is available at \url{https://github.com/codewhisperer-rs/IDP-DSN}.}
\end{abstract}

\ccsdesc[500]{Information systems~Data mining}
\ccsdesc[300]{Information systems~Social networks}

\keywords{Signed Networks, Dynamic Signed Networks,  Disentanglement}

\maketitle

\section{Introduction}
Signed networks are ubiquitous in the real world and describe both positive and negative relations among entities \cite{tang2016survey,leskovec2010chi}.
In online social platforms, such signed links often carry explicit, human-interpretable meanings: ``trust/support'' is usually positive while ``distrust/opposition'' is negative \cite{leskovec2010chi,leskovec2010predicting,guha2004propagation,ziegler2005propagation,dubois2011predicting}.
Beyond social systems, the same idea naturally extends to retrieval and recommender systems, where signed feedback can provide richer and more discriminative supervision than only clicks or ratings: positive links reinforce trust or preference, whereas negative links indicate rejection, distrust, or conflict \cite{massa2007trust,tang2016recommendations,victor2011trust,tang2015negative}.
Although negative feedback is often sparse in practice, prior studies show it can add unique value for robust modeling and recommendation \cite{kunegis2013added,tang2015negative,huang2023negative}.
Importantly, many modern platforms generate such signals as a time-stamped stream of interactions, so the underlying network continuously evolves with user activities and changing populations over time ~\cite{holme2012temporal}.
These observations naturally motivate learning on dynamic signed networks, where both topology and polarity change over time and should be interpreted jointly, rather than treated in isolation.

\begin{figure}[t]
  \centering
  \includegraphics[width=0.95\linewidth]{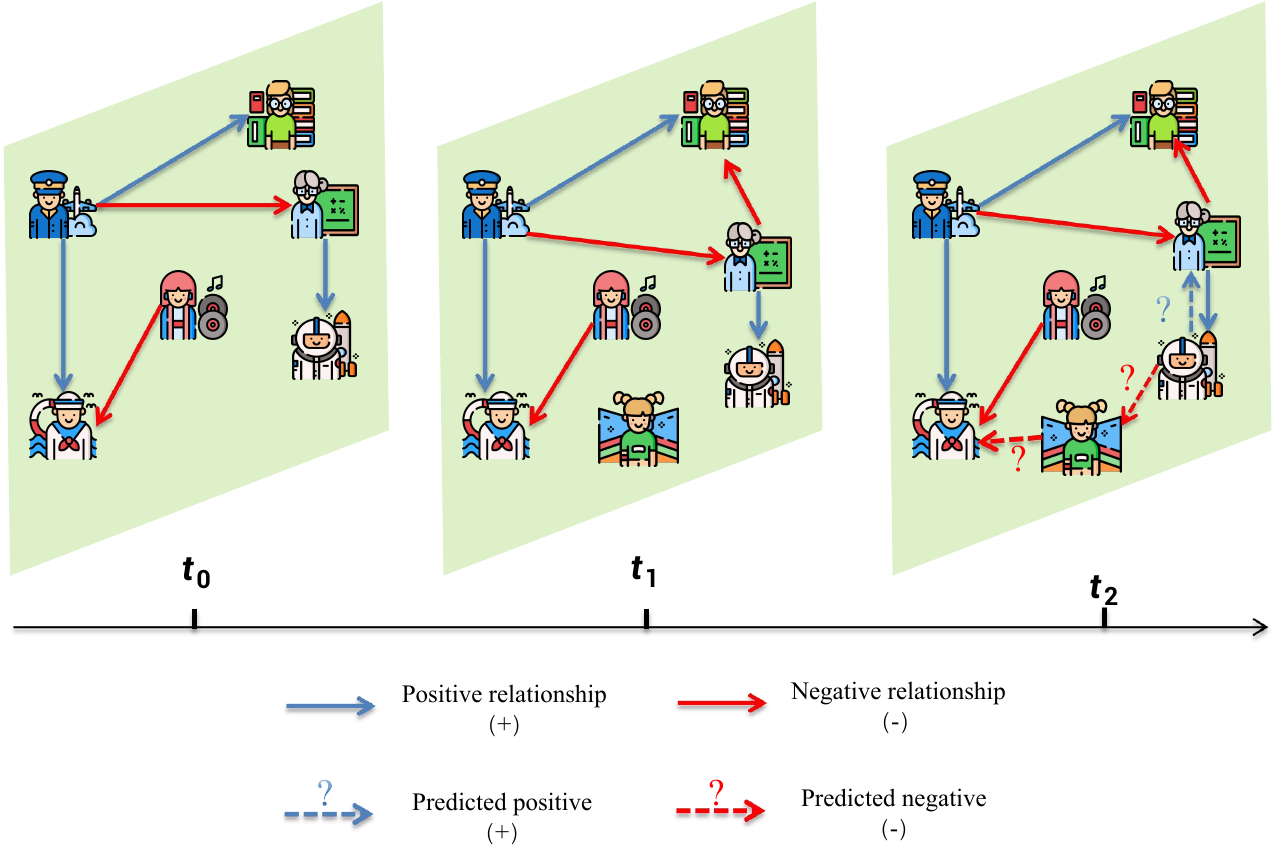}
  \caption{An illustration of dynamic signed networks for dynamic edge prediction.}
  \label{fig:figure1}
\end{figure}

In this work, we study dynamic edge prediction.
Given historical time-stamped signed interactions, the goal is to predict the future outcome for each candidate node pair as one of three classes: a non-existent edge, a positive edge, or a negative edge.
Here, a non-existent edge indicates that no interaction occurs between the pair in the prediction window.
This unified 3-way formulation aligns with practical decision-making and directly models both edge existence and polarity, avoiding brittle pipelines that separate them into disjoint subproblems.
Moreover, because the candidate space grows quadratically with the number of nodes, most node pairs correspond to non-existent edges, and negative edges are typically much rarer than positive ones, making the prediction problem highly imbalanced.
Compared with static link prediction~\cite{ran2024maximum}, dynamic prediction must reason about when and how relations evolve, and should exploit temporal dependencies rather than treating history as unordered evidence \cite{holme2012temporal,kumar2019predicting,sankar2020dysat}.
At the same time, real platforms face an inductive cold-start regime: new users/items arrive continuously and often provide only short and incomplete traces, making long node-specific trajectories unavailable \cite{hamilton2017inductive}.
As a result, models must generalize from limited observations and local signed-temporal structures, especially when test-time edges involve nodes that were not sufficiently observed during training.

Figure~\ref{fig:figure1} sketches our setting: signed interactions arrive as a time-stamped event stream, and the model is asked to predict whether a future relation will be positive, negative, or non-existent. Doing so requires capturing temporal recency and polarity-specific patterns, while remaining robust when only a small number of events are available for a node, which is common in inductive cold-start scenarios. Dynamic signed prediction is further complicated by two intrinsic properties of signed data. First, signed feedback is behaviorally and statistically asymmetric: negative events are rarer, yet they can carry disproportionate impact relative to positive evidence and may quickly overturn prior impressions \cite{baumeister2001bad,kunegis2013added,tang2015negative}. When both polarities are mixed into a single entangled temporal state, opposite signals can blur semantics or partially cancel each other, which is especially harmful under cold-start conditions with only a few observations. Second, from a structural perspective, signed networks are governed by global principles such as balance theory: positive links tend to cluster within groups while negative links often span group boundaries \cite{heider1946attitudes,cartwright1956structural,leskovec2010chi,traag2009community,esmailian2015signedcommunity,bonchi2019discovering}. As the network evolves, community boundaries and polarization states may shift, which can substantially change interaction outcomes \cite{antal2005dynamics,abell2009structural,marvel2011continuous}.

Dynamic signed network embedding has attracted increasing attention in recent years \cite{sun2024dynamise,sharma2023representation,kim2024polardsn}.
However, existing methods still miss important capabilities required by the above setting.
Many approaches for signed prediction or embedding are built on static snapshots and thus underuse timing information \cite{leskovec2010predicting,chiang2014prediction,derr2018signed}, while temporal graph models often update a single shared representation per node over time \cite{kumar2019predicting,sankar2020dysat}.
In signed networks, such coupling can entangle stable traits with short-term behaviors and mix positive/negative signals, weakening interpretability and harming inductive generalization when histories are short.
To address these challenges, we propose \textbf{IDP-DSN}, an inductive dual-polarity framework with polarity-wise static--dynamic disentanglement for dynamic signed networks.
IDP-DSN maintains sign-selective memories to capture polarity-asymmetric dynamics, performs history-only neighborhood inference for cold-start nodes (via sign-aware random walks, path encoding, and attention-based aggregation), and disentangles static identity from dynamic context within each polarity using an orthogonality regularizer; the dynamic update further incorporates a time-aware decay mechanism.
These designs reduce semantic interference, improve robustness under sparse histories, and better match polarity-asymmetric signed interactions.

Our main contributions are summarized as follows:
\begin{itemize}
  \item We propose \textbf{IDP-DSN}, an inductive dual-polarity framework that maintains sign-selective memories to model positive and negative temporal dynamics separately.
  \item We introduce polarity-wise static--dynamic disentanglement via an orthogonality regularizer, together with a time-aware decay mechanism for temporal attenuation.
  \item We design a history-only neighborhood inference mechanism for cold-start nodes and validate IDP-DSN on four real-world datasets (BitcoinAlpha, BitcoinOTC, Wiki-RfA, and Epinions), where it consistently achieves the best Macro-F1.
\end{itemize}

\section{Related Work}
In this section, we introduce some related works.
Our work relates to (i) \textbf{static signed network embeddings} (\textbf{SNE}),
(ii) \textbf{dynamic network embeddings} (\textbf{DNE}) for continuous-time event streams, and
(iii) \textbf{dynamic signed network embeddings} (\textbf{DSNE}).
This taxonomy also mirrors our experimental baselines, and highlights the gap we address: robust inductive modeling in dynamic signed networks under cold-start and sparse-history settings.

\subsection{Static Signed Network Embeddings (SNE)}
A fundamental challenge in signed networks is to represent antagonistic positive and negative links without collapsing them into a single proximity notion \cite{tang2016survey}.
Classic theories such as structural balance and its generalizations provide inductive biases for signed modeling \cite{heider1946attitudes,cartwright1956structural,davis1967clustering,doreian1996partitioning},
while signed directed networks further interact with status-related directionality \cite{leskovec2010chi}.

\textbf{Signed network embeddings.}
Based on these social theory (i.e., structural balance theory~\cite{heider1946attitudes} and status theory~\cite{leskovec2010chi}), 
early methods learn low-dimensional embeddings with sign-consistency constraints, e.g., SiNE \cite{wang2017signed} and SIDE for signed directed networks \cite{kim2018side}.
Spectral and Laplacian-based variants (e.g., SigMaNet) encode signed structure via magnetic or signed Laplacian operators \cite{fiorini2023sigmanet}.
These approaches are effective on snapshot tasks such as sign/link prediction \cite{leskovec2010predicting,chiang2014prediction}, but they treat the network as static.

\textbf{Signed GNNs.}
GNN-based approaches develop sign-aware message passing to separate evidence from opposite polarities~\cite{he2024pytorch,huang2021signed}, such as SGCN \cite{derr2018signed}, SiGAT \cite{huang2019signed}, and SDGNN \cite{huang2021sdgnn}.
Related variants study trustworthiness or latent grouping in signed graphs \cite{liu2021signed,kim2023trustsgcn}.
Despite strong performance, static SNE methods collapse time and thus cannot capture temporal recency, polarity drift, or evolving contexts in signed interactions \cite{tang2016survey,leskovec2010chi}.
This motivates event-driven modeling for signed networks.

\subsection{Dynamic Network Emebedings (DNE)}
\textbf{Dynamic network embeddings} (\textbf{DNE}) aims to learn node representations that evolve with time, capturing temporal recency and interaction dynamics.
A dominant paradigm is continuous-time temporal graph learning, which models interactions as time-stamped events and updates node states upon event arrivals.
Representative models include TGAT \cite{xu2020inductive}, TGN \cite{rossi2020temporal}, and DyRep \cite{trivedi2019dyrep}, with subsequent work improving efficiency and expressiveness (e.g., GraphMixer, DyFormer, TIGER) \cite{cong2023we,cong2023dyformer,zhang2023tiger}.

\textbf{Time encoding and temporal neighborhood aggregation.}
A common design is to encode time gaps or timestamps (e.g., sinusoidal or learned encodings) and combine them with historical neighborhoods via attention or message passing, enabling recency-aware modeling \cite{xu2020inductive,rossi2020temporal}.
Several works further improve inductive generalization beyond fixed node identities.
CAW summarizes historical structure via causal anonymous walks \cite{wang2021inductive}, and PINT proposes injective temporal message passing to better distinguish temporal neighborhoods \cite{souza2022provably}.

\textbf{Inductive setting and static--dynamic factorization.}
A key advantage of continuous-time models is inductive inference: representations can be computed from interaction histories rather than relying solely on node IDs \cite{xu2020inductive,rossi2020temporal}.
Another relevant thread decomposes node representations into time-invariant (stable identity/traits) and time-varying (short-term context/behavior) factors, which improves robustness under sparse observations and distribution shifts.
For example, DyTed separates static and dynamic components via adversarial objectives in dynamic graphs \cite{zhang2023dyted}.
However, most temporal graph models are developed for unsigned interactions; directly extending them to signed settings by treating polarity as a generic edge attribute can mix opposite-signed signals in a shared state,
leading to semantic mismatch or partial cancellation under polarity imbalance and sparse histories \cite{leskovec2010chi,tang2016survey}.
This motivates polarity-aware temporal mechanisms when moving to dynamic signed networks.

\subsection{Dynamic Signed Network Embeddings (DSNE)}
Dynamic signed network embeddings aim to jointly model temporal evolution and the antagonistic semantics of edge polarities.
Recent work includes DynamiSE \cite{sun2024dynamise}, SEMBA \cite{sharma2023representation}, and PolarDSN \cite{kim2024polardsn}, which also connect to studies of balance and polarization dynamics \cite{antal2005dynamics,abell2009structural,marvel2011continuous}.
Overall, DSNE methods suggest that accurate prediction in dynamic signed networks benefits from sign-aware temporal modeling.

Despite this progress, two challenges remain critical for our evaluation protocol.
First, polarity-specific dynamics are not always sufficiently isolated: mixing positive and negative evidence in shared updates can blur semantics or cancel signals, especially under sparse history \cite{sharma2023representation,kim2024polardsn}.
Second, cold-start inductive prediction is hard: emerging nodes have limited past events, and reliance on node-specific temporal trajectories can hinder generalization.
Our work bridges dynamic graph factorization and DSNE by explicitly disentangling static identity from dynamic context within each polarity and combining it with sign-aware temporal aggregation,
yielding a robust inductive model for dynamic signed networks under sparse-history and cold-start settings.

\section{Problem Definition}
\label{sec:Problem}

We study 3-class signed dynamic edge prediction on a dynamic signed network under a history protocol:
when predicting at time $t$, the model may only use interactions observed before $t$.
Accordingly, all representations and features at time $t$ are computed solely from the past event prefix $\mathcal{H}(t)$.

\textbf{Dynamic signed network.}
Let $\mathcal{V}$ be the node set and $\mathcal{E}=(e_1,\dots,e_{|\mathcal{E}|})$ be a time-ordered stream of interactions.
Each event is $e_k=(u_k,v_k,t_k,s_k,w_k)$, where $u_k$ and $v_k$ are the source and destination nodes,
$t_k\in\mathbb{R}_{>0}$ is the timestamp, $s_k\in\{+1,-1\}$ is the sign, and $w_k\in\mathbb{R}_{\ge 0}$ is the interaction weight.
For unweighted data, we set $w_k\equiv 1$ during preprocessing.
Events are ordered by non-decreasing timestamps; ties are broken by a fixed and dataset-consistent ordering (e.g., the listing order).
This convention is used only to define the information boundary and does not impose additional modeling assumptions.

\textbf{History and information boundary.}
For any time $t$, we define the available history as
\begin{equation}
\label{eq:hist_global}
\mathcal{H}(t)=\{e_i\in\mathcal{E}\mid t_i<t\}.
\end{equation}
When evaluating an event $e_k$ at time $t_k$, the model is allowed to use only $\mathcal{H}(t_k)$.
If multiple events share the same timestamp, we process them sequentially in the fixed tie-breaking order, treating them as
micro-steps; at each micro-step, the model may only access events from earlier micro-steps at that timestamp.

\textbf{3-class  dynamic signed edge prediction.}
We predict a label from
\begin{equation}
\label{eq:label_space}
\mathcal{Y}=\{\textsc{Pos},\textsc{Neg},\textsc{NonEdge}\}.
\end{equation}
A query instance is a tuple $(u,v,t)$, and the goal is to estimate the conditional distribution over $\mathcal{Y}$ given
the history $\mathcal{H}(t)$.
Let $y(u,v,t)\in\mathcal{Y}$ be defined as
\begin{equation}
\label{eq:label_def}
y(u,v,t)=
\begin{cases}
\textsc{Pos}, & \exists\, e_i=(u,v,t,+1,w_i)\in\mathcal{E},\\
\textsc{Neg}, & \exists\, e_i=(u,v,t,-1,w_i)\in\mathcal{E},\\
\textsc{NonEdge}, & \text{otherwise.}
\end{cases}
\end{equation}
In practice, \textsc{NonEdge} instances are generated by negative sampling at timestamps of observed edges.
These samples are used only as supervised training/evaluation instances and are not inserted into the event stream;
therefore, \textsc{NonEdge} samples never contribute to $\mathcal{H}(t)$.

The learning objective is to estimate a predictor
\begin{equation}
\label{eq:pred_def}
f_\theta:\ (u,v,t,\mathcal{H}(t))\ \mapsto\ p_\theta(y\mid u,v,t)\in\Delta^{|\mathcal{Y}|},
\end{equation}
where $f_\theta$ is constrained to depend on the history $\mathcal{H}(t)$ at time $t$.

\textbf{Inductive setting.}
In inductive evaluation, some nodes that appear during testing are unseen in training.
The predictor must handle interactions involving such nodes without retraining, using the learned parameters and the history that becomes available as the stream progresses.

\section{Methodology} 
\label{sec:method}

We propose \textbf{IDP-DSN} for \textbf{3-class signed dynamic edge prediction} with
$\mathcal{Y}=\{\textsc{Pos},\textsc{Neg},\textsc{NonEdge}\}$.
Given a query $(u,v,t)$, all representations are computed from the available history $\mathcal{H}(t)$
(Section~\ref{sec:Problem}), with micro-step processing for timestamp ties.
\textsc{NonEdge} instances are generated by negative sampling at observed-edge timestamps and are not inserted into the
event stream, thus they never contribute to $\mathcal{H}(t)$.

\begin{figure*}[t]
  \centering
  \includegraphics[width=\textwidth]{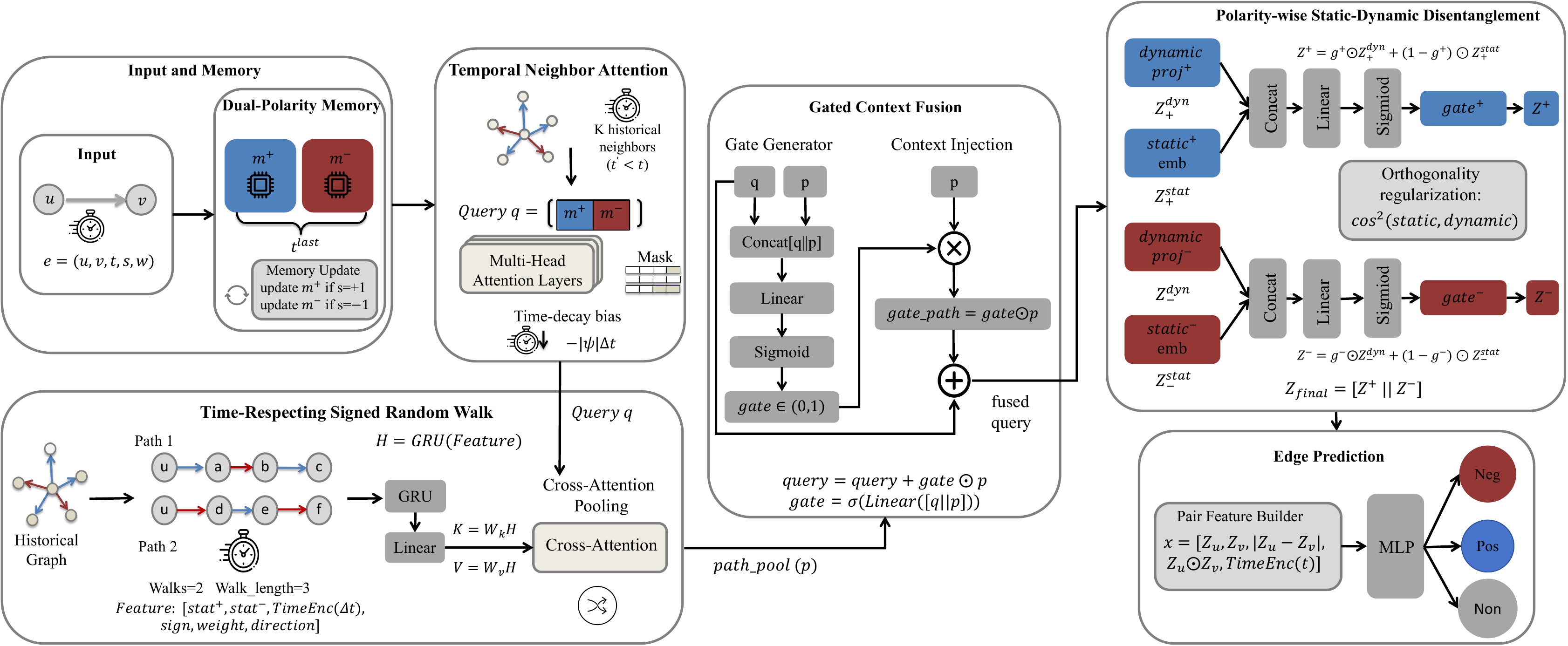}
  \caption{Overall framework of \textbf{IDP-DSN} for 3-class dynamic edge prediction.
  Given a query $(u,v,t)$, the model reads dual-polarity memories, aggregates strictly causal neighbors via time-decay attention,
  and encodes strictly causal random-walk paths with a GRU.
  Cross-attention pools a path context vector $\mathbf{p}$, which is injected by gated residual fusion.
  Polarity-wise projection and fusion yield representations for \textsc{Pos}/\textsc{Neg}/\textsc{NonEdge}.
  The sign of the current observation is used only for supervision and post-event memory update (never as a forward input).}
  \label{fig:framework}
\end{figure*}

Figure~\ref{fig:framework} overviews IDP-DSN: it combines dual-polarity memory, time-decay neighbor attention, and a
walk-based context encoder with gated fusion, followed by polarity-wise static--dynamic disentanglement for 3-class prediction.
For an observed event, prediction is made before memory update, so the current sign is never used as a forward input.

\subsection{Dual-Polarity Memory}
\label{subsec:memory}
Each node $u$ maintains two polarity-specific temporal memories
$\mathbf{m}_u^{+}(t),\mathbf{m}_u^{-}(t)\in\mathbb{R}^{d_{emb}}$ and a last-update time $t_u^{last}$.

\textbf{Coupled message construction.}
After an observed event $e=(u,v,t,s,w)$ is revealed at its micro-step, we construct an update message using both endpoints'
memories immediately before the update:
\begin{equation}
\label{eq:msg}
\begin{split}
\mathbf{msg}_{uv}(t)
&=\mathrm{MLP}_{msg}\!\Big(
\mathbf{m}_u^{+}(t^-)\Vert \mathbf{m}_u^{-}(t^-)\Vert{}\\
&\qquad \mathbf{m}_v^{+}(t^-)\Vert \mathbf{m}_v^{-}(t^-)\Vert
w\Vert s
\Big),
\end{split}
\end{equation}
where $t^-$ denotes the memory state immediately before processing the current observation under the fixed micro-step ordering, $w$ is the observed scalar interaction weight set to $1$ for unweighted data, and $s\in\{+1,-1\}$ is the sign of the observed event. The sign $s$ is used only for the post-event memory update and supervision, and is not a forward input when computing the prediction at time $t$.

\textbf{Sign-specific memory update.}
Let $\tilde{\Delta t}_u=\log(1+\max(t-t_u^{last},0))$ denote the elapsed time since $u$ was last updated, which provides
temporal context for irregularly-timed interaction streams. For $s\in\{+1,-1\}$, we update only the matching channel:
\begin{equation} \label{eq:mem_update}
\begin{aligned}
\mathbf{m}_u^{s}(t) &=\mathrm{GRU}_s\!\Big(\big[\mathbf{msg}_{uv}(t)\Vert \tilde{\Delta t}_u\big],\,\mathbf{m}_u^{s}(t^-)\Big),\\
\mathbf{m}_u^{-s}(t) &=\mathbf{m}_u^{-s}(t^-).
\end{aligned}
\end{equation}
We update $v$ in the same way. After processing the event at time $t$, we record the last-update times as
$t_u^{last}=t$ and $t_v^{last}=t$.
\textsc{NonEdge} samples never trigger memory updates.

\subsection{Temporal Neighbor Attention} \label{subsec:neighbor}
For a query $(u,v,t)$, IDP-DSN computes endpoint representations separately; we describe the construction for $u$.

\textbf{Query representation.}
We form the query representation of node $u$ by concatenating its dual-polarity memories read at time $t$:
\[
\mathbf{q}=\big[\mathbf{m}_u^{+}(t)\Vert \mathbf{m}_u^{-}(t)\big]\in\mathbb{R}^{2d_{emb}},
\]
where $\mathbf{m}_u^{+}(t)\in\mathbb{R}^{d_{emb}}$ and $\mathbf{m}_u^{-}(t)\in\mathbb{R}^{d_{emb}}$ denote the positive and
negative memories of node $u$ computed from the available history $\mathcal{H}(t)$.

\textbf{Historical neighbors lookup from $\mathcal{H}(t)$.}
We collect up to $K$ historical interactions in $\mathcal{H}(t)$ that involve node $u$.
Each retrieved interaction is either an outgoing event $e_k=(u,v_k,t_k,s_k,w_k)$ or an incoming event
$e_k=(v_k,u,t_k,s_k,w_k)$ with $t_k<t$, where $v_k$ is the neighbor node of $u$ in this interaction.

\textbf{Learnable time encoding.}
We encode a scalar time lag $\tau$ as
\[
\mathrm{TimeEnc}(\tau)=\mathrm{MLP}_{time}([\tau])\in\mathbb{R}^{d_{time}},
\]
where $[\tau]$ denotes the 1-D input and $d_{time}$ is the time-embedding dimension.

\textbf{Neighbors feature construction.}
For each retrieved interaction with neighbor $v_k$, we construct a neighbor feature vector by concatenating the neighbor's
dual-polarity memories, the learnable time encoding, and two scalar attributes:
\begin{equation}\label{eq:neighbor_token}
\mathbf{h}_k=
\Big[
\mathbf{m}_{v_k}^{+}(t)\Vert \mathbf{m}_{v_k}^{-}(t)\Vert
\mathrm{TimeEnc}(t-t_k)\Vert s_k\Vert w_k
\Big],
\end{equation}
where $\mathbf{m}_{v_k}^{+}(t)\in\mathbb{R}^{d_{emb}}$ and $\mathbf{m}_{v_k}^{-}(t)\in\mathbb{R}^{d_{emb}}$ are the
positive and negative memories of $v_k$ read at time $t$, $\mathrm{TimeEnc}(t-t_k)\in\mathbb{R}^{d_{time}}$ is the time
encoding of the lag $t-t_k$, $s_k\in\{+1,-1\}$ is the sign of the $k$-th historical interaction, and
$w_k\in\mathbb{R}_{\ge 0}$ is its scalar weight. For unweighted data, we set $w_k=1$ during preprocessing.
Both $s_k$ and $w_k$ are concatenated as scalars without an embedding layer. Therefore,
\begin{equation}\label{eq:neighbor_dim}
\dim(\mathbf{h}_k)=2d_{emb}+d_{time}+3.
\end{equation}
We apply an attention mask so that only the retrieved historical interactions are visible to the attention computation.

\textbf{Time-decay multi-head attention.}
We apply $N$ stacked multi-head attention layers with residual connections and feed-forward blocks.
For head $h$, the attention weight on neighbor $k$ is
\begin{equation} \label{eq:attn}
\alpha_k^{(h)}=
\mathrm{Softmax}_k\!\Bigg(
\frac{\langle \mathbf{W}_q^{(h)}\mathbf{q},\,\mathbf{W}_k^{(h)}\mathbf{h}_k\rangle}{\sqrt{d_h}}
-|\psi_h|\,(t-t_k)
\Bigg),
\end{equation}
where $\psi_h$ is a learnable decay rate and $d_h$ is the per-head dimension. The updated query representation is computed by standard attention aggregation and then passed through the feed-forward block; after $N$ layers, we obtain the
history-aware query representation, denoted again by $\mathbf{q}$.

\subsection{Random Walk Aggregation}
\label{subsec:walk}
Temporal neighbors provide primarily 1-hop evidence. We further extract higher-order signed structural signals by sampling random walks on the historical graph induced by $\mathcal{H}(t)$, following classical random-walk based representation learning\cite{perozzi2014deepwalk,grover2016node2vec}.

\textbf{Time-decay biased walk sampling.}
Let $\mathcal{G}(t)$ denote the graph constructed from $\mathcal{H}(t)$.
Starting from $u$, we sample $R$ random walks of length $L$ on $\mathcal{G}(t)$.
At each step, we collect as candidates all historical interactions in $\mathcal{H}(t)$ that are incident to the current
node. Among these candidates, we sample the next hop using a time-decay biased distribution, so that interactions closer to
the query time are more likely to be selected.
Specifically, for a candidate interaction with timestamp $t_i$, we define $\Delta t_i=t-t_i$ and assign an unnormalized
weight
\begin{equation} \label{eq:rw_decay}
\pi_i \propto \exp(-\gamma\,\Delta t_i),
\end{equation}
where $\gamma>0$ is a decay hyperparameter; probabilities are obtained by normalizing $\{\pi_i\}$ over the candidate set at
the current step. This sampling neither separates positive and negative edges nor uses interaction weights for transition;
the bias is solely determined by temporal proximity in Eq.~\eqref{eq:rw_decay}.

\textbf{Walk step features.}
For a walk step visiting node $v_i$ through a historical interaction with timestamp $t_i$, sign $s_i$, weight $w_i$, and
direction indicator $d_i$, we build
\begin{equation}\label{eq:step_feat}
\mathbf{x}^{(i)}_{step}=
\Big[
\mathbf{z}_{v_i,+}^{stat}\Vert \mathbf{z}_{v_i,-}^{stat}\Vert
\mathrm{TimeEnc}(t-t_i)\Vert s_i\Vert w_i\Vert d_i
\Big],
\end{equation}
where $\mathbf{z}^{stat}_{v_i,+}$ and $\mathbf{z}^{stat}_{v_i,-}$ are polarity-wise static embeddings of node $v_i$ learned as lookup-table parameters, and $\mathrm{TimeEnc}(t-t_i)$ is the learnable encoding of the lag $t-t_i$.

\textbf{GRU encoding and cross-attention pooling.}
A GRU encodes each sampled walk sequence and produces hidden states $H$.
We linearly project hidden states to keys and values as $K=\mathbf{W}_K H$ and $V=\mathbf{W}_V H$.
We then pool a walk representation by cross-attention using the neighbor-aware query representation $\mathbf{q}$:
\begin{equation} \label{eq:path_pool}
\mathbf{p}=\textit{path\_pool}=\mathrm{CrossAttn}(\mathbf{q},K,V),
\end{equation}
where $\mathrm{CrossAttn}(\mathbf{q},K,V)$ is standard multi-head scaled dot-product cross-attention with $\mathbf{q}$ as the
query and the GRU-derived sequences $K$ and $V$ as keys and values.

\subsection{Gated Context Fusion} 
\label{subsec:fusion}
We fuse the walk-derived representation $\mathbf{p}$ into the neighbor-aware query representation $\mathbf{q}$ through a gated residual connection. This design allows the model to adaptively control the contribution of walk-based signals,so that $\mathbf{p}$ is emphasized when it provides complementary higher-order evidence and suppressed when local neighbor
evidence is sufficient. The residual form preserves the original neighbor-aware representation, while the gate provides a data-dependent mechanism to modulate the injected walk signal.

\textbf{Gate generation and context injection.}
We compute a gate from the concatenation of $\mathbf{q}$ and $\mathbf{p}$ and use it to modulate $\mathbf{p}$ before adding
it back to the query:
\begin{equation} 
\label{eq:gated_fusion}
\begin{aligned}
\mathbf{g} &= \sigma\!\left(\mathbf{W}_{g}[\mathbf{q}\Vert \mathbf{p}]\right),\\
\mathbf{q}' &= \mathbf{q} + \mathbf{g}\odot \mathbf{p},
\end{aligned}
\end{equation}
where $\sigma(\cdot)$ is the sigmoid function and $\odot$ denotes element-wise product.
The gate $\mathbf{g}\in(0,1)^d$ performs dimension-wise selection. We use $\mathbf{q}'$ as the fused query representation and pass it to the subsequent polarity-wise projection and fusion module.

\subsection{Polarity-wise Static--Dynamic Disentanglement} 
\label{subsec:disent}
Given the fused query representation $\mathbf{q}$, we compute polarity-wise dynamic factors and combine them with polarity-wise static embeddings.

\textbf{Dynamic projection.}
\begin{equation} \label{eq:dyn_proj}
\mathbf{z}^{dyn}_{u,+}(t)=\mathrm{Proj}_{+}(\mathbf{q}),\qquad
\mathbf{z}^{dyn}_{u,-}(t)=\mathrm{Proj}_{-}(\mathbf{q}),
\end{equation}
where $\mathrm{Proj}_{\pm}(\cdot)$ are learnable projection heads.

\textbf{Static embeddings.}
We maintain polarity-specific embedding tables to provide static identity features:
\begin{equation} \label{eq:stat_emb}
\mathbf{z}^{stat}_{u,+}=\mathrm{Emb}_{+}(u),\qquad
\mathbf{z}^{stat}_{u,-}=\mathrm{Emb}_{-}(u).
\end{equation}
For node identifiers that are unseen during training in inductive evaluation, $\mathrm{Emb}_{+}(\cdot)$ and
$\mathrm{Emb}_{-}(\cdot)$ return a shared learnable unknown-node embedding.

\textbf{Polarity-wise gated fusion.}
We fuse static and dynamic components for the positive and negative channels with learnable gates:
\begin{equation} \label{eq:pol_gate_pos}
\mathbf{g}_{u,+}(t)=\sigma\!\left(\mathbf{W}_{g,+}\big[\mathbf{z}^{dyn}_{u,+}(t)\Vert \mathbf{z}^{stat}_{u,+}\big]\right),
\end{equation}
\begin{equation} \label{eq:pol_fusion_pos}
\mathbf{z}^{final}_{u,+}(t)=
\mathbf{g}_{u,+}(t)\odot \mathbf{z}^{dyn}_{u,+}(t)+
\big(1-\mathbf{g}_{u,+}(t)\big)\odot \mathbf{z}^{stat}_{u,+},
\end{equation}
\begin{equation} \label{eq:pol_gate_neg}
\mathbf{g}_{u,-}(t)=\sigma\!\left(\mathbf{W}_{g,-}\big[\mathbf{z}^{dyn}_{u,-}(t)\Vert \mathbf{z}^{stat}_{u,-}\big]\right),
\end{equation}
\begin{equation} \label{eq:pol_fusion_neg}
\mathbf{z}^{final}_{u,-}(t)=
\mathbf{g}_{u,-}(t)\odot \mathbf{z}^{dyn}_{u,-}(t)+
\big(1-\mathbf{g}_{u,-}(t)\big)\odot \mathbf{z}^{stat}_{u,-}.
\end{equation}
We concatenate 
$\mathbf{z}^{final}_{u}(t)=\big[\mathbf{z}^{final}_{u,+}(t)\Vert \mathbf{z}^{final}_{u,-}(t)\big]$.

\textbf{Orthogonality regularization.}
We encourage static and dynamic components to carry complementary information via squared cosine similarity:
\begin{equation} \label{eq:orth}
\begin{aligned}
\mathcal{L}_{orth}
&=\frac{1}{2|\mathcal{B}|}\sum_{u\in\mathcal{B}}
\Big[
\cos^2(\mathbf{z}^{stat}_{u,+},\mathbf{z}^{dyn}_{u,+}(t)) \\
&\qquad\qquad
+\cos^2(\mathbf{z}^{stat}_{u,-},\mathbf{z}^{dyn}_{u,-}(t))
\Big],
\end{aligned}
\end{equation}
where $\mathcal{B}$ denotes the mini-batch node set, $|\mathcal{B}|$ is its cardinality, and $\cos(\cdot,\cdot)$ is cosine similarity.

\subsection{Edge Prediction} 
\label{subsec:predhead}
Given a query $(u,v,t)$, we first obtain the endpoint representations $\mathbf{z}^{final}_{u}(t)$ and
$\mathbf{z}^{final}_{v}(t)$ by applying the above modules to each endpoint using only the available history
$\mathcal{H}(t)$. We then construct a symmetric pairwise feature vector that combines node-level information and
order-invariant interaction features:
\begin{equation}
\label{eq:pairfeat}
\begin{split}
\mathbf{f}_{pair}(u,v,t)=\Big[
\mathbf{z}^{final}_u(t)\Vert \mathbf{z}^{final}_v(t)\Vert{}\\
\qquad \mathbf{z}^{diff}_{uv}(t)\Vert \mathbf{z}^{prod}_{uv}(t)\Vert
\mathrm{TimeEnc}(t)
\Big],
\end{split}
\end{equation}
where $\mathbf{z}^{diff}_{uv}(t)$ captures absolute differences between endpoints and $\mathbf{z}^{prod}_{uv}(t)$ captures
feature-wise compatibility through element-wise products:
\begin{equation}
\label{eq:pairfeat_aux}
\begin{split}
\mathbf{z}^{diff}_{uv}(t)&=\big|\mathbf{z}^{final}_u(t)-\mathbf{z}^{final}_v(t)\big|,\\
\mathbf{z}^{prod}_{uv}(t)&=\mathbf{z}^{final}_u(t)\odot \mathbf{z}^{final}_v(t).
\end{split}
\end{equation}
We append $\mathrm{TimeEnc}(t)$ to provide a learnable representation of the query time.
Finally, a multi-layer perceptron followed by a softmax produces a probability distribution over the three labels:
\begin{equation} \label{eq:pred}
P_\theta(y\mid u,v,t)=\mathrm{Softmax}\!\left(\mathrm{MLP}_\theta\!\left(\mathbf{f}_{pair}(u,v,t)\right)\right),
\quad y\in\mathcal{Y}.
\end{equation}

\subsection{Training Objective} 
\label{subsec:train}

\textbf{\textsc{NonEdge} examples.}
We construct \textsc{NonEdge} instances to train the 3-class predictor.
For each observed event $e_k=(u_k,v_k,t_k,s_k,w_k)$, we sample node pairs $(u',v')$ at the same timestamp $t_k$ and label them as \textsc{NonEdge} when no interaction occurs between $u'$ and $v'$ at time $t_k$.
These samples are used only as supervised instances and are not inserted into the event stream $\mathcal{E}$; therefore, they never contribute to $\mathcal{H}(t)$ and never trigger memory updates.

\textbf{Class reweighting and objective.}
We apply batch-wise square-root tempered inverse-frequency weighting. For class $c\in\mathcal{Y}$ with batch count $N_c$
and $N=\sum_{c'\in\mathcal{Y}}N_{c'}$:
\begin{equation} \label{eq:weights}
\omega_c= \frac{\sqrt{N/(N_c+\epsilon)}}{\frac{1}{|\mathcal{Y}|}\sum_{c'\in\mathcal{Y}}\sqrt{N/(N_{c'}+\epsilon)}},
\qquad |\mathcal{Y}|=3,
\end{equation}
where $\epsilon>0$ is a small constant for numerical stability and the normalization ensures that the average weight across
the three classes equals $1$ within each batch.
We minimize the weighted cross-entropy:
\begin{equation} \label{eq:taskloss}
\mathcal{L}_{task}=-\sum_{e\in\mathcal{B}} \omega_{y_e}\log P_\theta(y_e\mid u,v,t),
\end{equation}
and the full objective is
\begin{equation} \label{eq:totloss}
\mathcal{L}=\mathcal{L}_{task}+\lambda \mathcal{L}_{orth},
\end{equation}
where $\lambda\ge 0$ controls the strength of orthogonality regularization.

\section{Experiments}

We evaluate IDP-DSN on four real-world signed temporal networks and study:
(i) overall effectiveness on 3-class dynamic edge prediction,
(ii) contributions of key components, and
(iii) sensitivity to major hyperparameters.
We report results on the \textbf{Hybrid} test set and further break down performance into \textbf{Transductive} and \textbf{Inductive} subsets.

We want to answer the following Research Questions (RQs):
\begin{itemize}
  \item \textbf{RQ1:} Does IDP-DSN outperform strong baselines on dynamic edge prediction, formulated as a three-way classification over non-edge, positive, and negative outcomes?
  \item \textbf{RQ2:} How does performance change when we ablate core components of IDP-DSN, including removing polarity separation in Dual-Polarity Memory, disabling Time-decay in Temporal Neighbor Attention, removing Random Walk Aggregation and its cross-attention pooling, and replacing the full model with static-only(w/o Dynamic) or dynamic-only(w/o Static) variants?
    
  \item \textbf{RQ3:} How sensitive is IDP-DSN to the embedding dimension $d$, orthogonality weight $\lambda$, walk length $L$, and the number of sampled walks $D$?
\end{itemize}

\begin{table}
  \centering
  \caption{Statistics of the signed temporal network datasets. \#Timestamps is the number of unique time indices; Pos(\%)/Neg(\%) are class proportions; Time Span is the temporal coverage.}

  \label{tab:dataset-stats}
  \scriptsize
  \setlength{\tabcolsep}{3.5pt} %
  \renewcommand{\arraystretch}{1.05}
  \resizebox{\linewidth}{!}{%
    \begin{tabular}{lrrrrrr}
      \toprule
      Dataset & Nodes & Edges & \#Timestamps & Pos (\%) & Neg (\%) & Time Span \\
      \midrule
      BitcoinAlpha & 3{,}783   & 24{,}186   & 1{,}647   & 93.6 & 6.4  & $\sim$5 years \\
      BitcoinOTC   & 5{,}881   & 35{,}592   & 35{,}592  & 90.0 & 10.0 & $\sim$5 years \\
      Wiki-RfA     & 10{,}595  & 170{,}499  & 162{,}274 & 77.5 & 22.5 & $\sim$10 years \\
      Epinions     & 131{,}828 & 841{,}372  & 939       & 85.3 & 14.7 & $\sim$2 years \\
      \bottomrule
    \end{tabular}%
  }
\end{table}

\begin{table*}[!t]
\centering
\caption{Dynamic edge prediction results on the \textbf{Hybrid} test set. We compare 13 methods and report Acc., W-F1, and M-F1; higher is better. Bold denotes the best and underline denotes the second best in each column.}
\label{tab:overall}
\begin{adjustbox}{width=\textwidth}
\begin{tabular}{l ccc ccc ccc ccc}
\toprule
\multirow{2}{*}{\raisebox{-0.6ex}{Method}}
& \multicolumn{3}{c}{BitcoinAlpha}
& \multicolumn{3}{c}{BitcoinOTC}
& \multicolumn{3}{c}{Wiki-RfA}
& \multicolumn{3}{c}{Epinions} \\
\cmidrule(lr){2-4}\cmidrule(lr){5-7}\cmidrule(lr){8-10}\cmidrule(lr){11-13}
& Acc. & W-F1 & M-F1
& Acc. & W-F1 & M-F1
& Acc. & W-F1 & M-F1
& Acc. & W-F1 & M-F1 \\
\midrule
\multicolumn{13}{>{\columncolor{gray!15}}c}{\textbf{DNE}} \\
TGAT       & 0.7223 & 0.6958 & 0.5019 & 0.7488 & 0.7370 & 0.5509 & 0.6464 & 0.6479 & 0.5476 & 0.6337 & 0.5941 & 0.4510 \\
TGN        & 0.7053 & 0.6790 & 0.4888 & 0.7042 & 0.6772 & 0.4836 & 0.6146 & 0.5809 & 0.4380 & 0.7160 & 0.6871 & 0.5249 \\
CAW        & 0.7035 & 0.6745 & 0.4857 & 0.7234 & 0.6953 & 0.4970 & 0.6510 & 0.6063 & 0.4519 & \multicolumn{3}{c}{out of memory} \\
PINT       & 0.5412 & 0.5160 & 0.3851 & 0.6039 & 0.5702 & 0.4080 & 0.5413 & 0.4660 & 0.3368 & \multicolumn{3}{c}{out of memory} \\
GraphMixer & 0.7321 & 0.7222 & 0.5456 & 0.6691 & 0.6904 & 0.5483 & 0.7162 & 0.6983 & 0.5700 & 0.7584 & 0.7560 & 0.6797 \\
TIGER      & 0.7202 & 0.7242 & 0.5684 & 0.6023 & 0.6194 & 0.5026 & 0.7160 & 0.7103 & 0.5786 & 0.6781 & 0.6863 & 0.6007 \\
DyFormer   & 0.7083 & 0.6806 & 0.4909 & 0.7102 & 0.6879 & 0.5040 & 0.6019 & 0.5632 & 0.4238 & 0.7340 & 0.7168 & 0.5953 \\
\midrule
\multicolumn{13}{>{\columncolor{gray!15}}c}{\textbf{SNE}} \\
SGCN     & 0.4901 & 0.4341 & 0.3162 & 0.5579 & 0.5135 & 0.3680 & 0.4698 & 0.3785 & 0.2682 & 0.6478 & 0.6103 & 0.4770 \\
SIGAT    & 0.5343 & 0.4544 & 0.3229 & 0.6055 & 0.5635 & 0.4174 & 0.4326 & 0.3617 & 0.2707 & 0.6996 & 0.6791 & 0.5804 \\
SDGNN    & 0.5941 & 0.5540 & 0.4081 & 0.6253 & 0.5758 & 0.4102 & 0.4833 & 0.3962 & 0.2888 & 0.6863 & 0.6591 & 0.5376 \\
SigMaNet & 0.5476 & 0.4739 & 0.3386 & 0.6428 & 0.6128 & 0.4619 & 0.5214 & 0.4603 & 0.3487 & 0.6538 & 0.6255 & 0.5310 \\
\midrule
\multicolumn{13}{>{\columncolor{gray!15}}c}{\textbf{DSNE}} \\
SEMBA    & 0.7126 & 0.7234 & 0.6618 & 0.7377 & 0.7473 & 0.6786 & 0.6203 & 0.6616 & 0.5985 & 0.6703 & 0.6666 & 0.6380 \\
PolarDSN & \underline{0.8082} & \underline{0.8155} & \underline{0.6922} & \underline{0.8365} & \underline{0.8539} & \underline{0.7115} & \underline{0.8319} & \underline{0.8443} & \underline{0.7162} & \underline{0.7862} & \underline{0.7993} & \underline{0.7148} \\
IDP-DSN(ours) & \textbf{0.8705} & \textbf{0.8682} & \textbf{0.8475} & \textbf{0.9223} & \textbf{0.9207} & \textbf{0.8743} & \textbf{0.9452} & \textbf{0.9447} & \textbf{0.8989} & \textbf{0.9307} & \textbf{0.9302} & \textbf{0.9100} \\
\bottomrule
\end{tabular}
\end{adjustbox}
\end{table*}

\subsection{Experimental Setup}
We describe the datasets, evaluation protocol, baselines, and implementation details.To ensure fair comparison and reproducibility, we use a strict chronological split and reuse the same pre-generated evaluation negatives for every dataset.

\subsubsection{Datasets}
We evaluate on four real-world signed temporal networks: BitcoinAlpha, BitcoinOTC, Wiki-RfA, and Epinions.
Each dataset consists of timestamped directed signed edges, where edge polarities represent trust/distrust or support/oppose relations.For the observed signed interaction stream (\textsc{Pos}/\textsc{Neg}), we directly use the preprocessed event data released by PolarDSN \cite{kim2024polardsn} and do not apply any additional filtering or transformation.
Our only additional processing is the construction of evaluation \textsc{NonEdge} pairs, as described below.
Table~\ref{tab:dataset-stats} summarizes the dataset statistics.
These benchmarks are widely used in prior studies on signed networks.
\cite{derr2018signed,fiorini2023sigmanet,huang2019signed,huang2021sdgnn,kim2023trustsgcn,liu2021signed,sharma2023representation}.

\subsubsection{Evaluation Protocol}
We evaluate dynamic edge prediction as a three class classification task with
$\mathcal{Y}=\{\text{\textsc{pos}}, \text{\textsc{neg}}, \text{\textsc{nonedge}}\}$.
Following PolarDSN~\cite{kim2024polardsn}, we adopt a chronological and node aware protocol to jointly assess temporal generalization and inductive performance.
The full protocol is summarized below.

\begin{itemize}
  \item \textbf{Dataset split.}
  For each dataset, interactions are split by timestamp into training (70\%), validation (15\%), and testing (15\%).
  The training set contains edges in the first 70\% time quantile, the validation set lies in the 70\% to 85\% window, and the test set contains edges after the 85\% quantile.
  This chronological split prevents future information from leaking into model fitting and evaluates temporal generalization under a realistic deployment order.

  \item \textbf{Masked nodes for inductive evaluation.}
  To evaluate a cold start regime, we mask a subset of nodes so that they are unseen during training, following PolarDSN~\cite{kim2024polardsn}.
  We first form a candidate pool as the union of all source and destination nodes that appear in interactions after the validation cutoff time, namely the 70\% time quantile.
  From this pool, we uniformly sample masked nodes, with the sample size set to 10\% of the total number of unique nodes in the full dataset.
  To ensure that these masked nodes are truly unseen in training, we remove from the training window any edge incident to a masked node.

  \item \textbf{Transductive, inductive, and hybrid test subsets.}
  Under this protocol, we report results on three evaluation subsets derived from the test window.
  The transductive subset contains test edges whose two endpoints are both observed in training.
  The inductive subset contains test edges where at least one endpoint is a masked node.
  The hybrid subset corresponds to the full test set.
  In the main paper, we report hybrid performance and also provide transductive and inductive results for a more fine grained analysis.

 \item \textbf{Evaluation \textsc{NonEdge} sampling.}
During training, \textsc{NonEdge} pairs are sampled on the fly.
For validation and test, we pre-generate the negative pairs once and reuse them across all methods and runs to eliminate evaluation randomness.
For each positive interaction $(u,v,t)$, we generate one negative pair $(u',v',t)$ (a $1{:}1$ ratio), keeping the timestamp identical to the positive sample.
Both endpoints are sampled independently and uniformly from the evaluation-time node set $\mathcal{V}_{\text{eval}}$:
\begin{equation}
u' \sim \mathcal{U}(\mathcal{V}_{\text{eval}}), \qquad v' \sim \mathcal{U}(\mathcal{V}_{\text{eval}}).
\end{equation}
We do not enforce rejection sampling to exclude true edges, as collisions are negligible due to the extreme sparsity of the interaction graphs.
The resulting evaluation negatives are cached and kept unchanged for the remainder of the experiments.

\end{itemize}

\subsubsection{Metrics}
We report Accuracy, Weighted-F1, and Macro-F1. Macro-F1 is our primary metric because it averages the per-class F1 scores of \textsc{Pos}, \textsc{Neg}, and \textsc{Nonedge}, and is robust to class imbalance (especially the scarcity of \textsc{Neg} edges). Weighted-F1 weights each class by its frequency in the evaluation set and reflects performance under the empirical distribution.

\begin{table}[!t]
\centering
\caption{Macro-F1 results for dynamic edge prediction on \textbf{Transductive} and \textbf{Inductive} test edges. Higher is better. Bold denotes the best and underline denotes the second best in each column.}
\label{tab:trans-induc}
\scriptsize
\setlength{\tabcolsep}{3.5pt}
\renewcommand{\arraystretch}{1.1}
\begin{adjustbox}{width=\columnwidth,center}
\begin{tabular}{l cc cc cc cc}
\toprule
\multirow{2}{*}{Method}
& \multicolumn{2}{c}{BitcoinAlpha}
& \multicolumn{2}{c}{BitcoinOTC}
& \multicolumn{2}{c}{Wiki-RfA}
& \multicolumn{2}{c}{Epinions} \\
\cmidrule(lr){2-3}\cmidrule(lr){4-5}\cmidrule(lr){6-7}\cmidrule(lr){8-9}
& Trans. & Induc. & Trans. & Induc. & Trans. & Induc. & Trans. & Induc. \\
\midrule
\multicolumn{9}{>{\columncolor{gray!15}}c}{\textbf{DNE}} \\
TGAT       & 0.4994 & 0.4978 & 0.5899 & 0.5412 & 0.5100 & 0.5494 & 0.6094 & 0.4166 \\
TGN        & 0.4935 & 0.4785 & 0.4512 & 0.4581 & 0.3505 & 0.4356 & 0.5535 & 0.5253 \\
CAW        & 0.5349 & 0.4724 & 0.5795 & 0.4732 & 0.5021 & 0.4486 & \multicolumn{2}{c}{out of memory} \\
PINT       & 0.3614 & 0.4221 & 0.4478 & 0.4164 & 0.2702 & 0.3562 & \multicolumn{2}{c}{out of memory} \\
GraphMixer & 0.5093 & 0.5340 & 0.4785 & 0.5453 & 0.3670 & 0.5699 & 0.6491 & 0.6828 \\
TIGER      & 0.5914 & 0.5539 & 0.5640 & 0.5045 & 0.5550 & 0.5935 & 0.6284 & 0.6275 \\
DyFormer   & 0.5253 & 0.4852 & 0.5647 & 0.4876 & 0.4826 & 0.4161 & 0.6107 & 0.5831 \\
\midrule
\multicolumn{9}{>{\columncolor{gray!15}}c}{\textbf{SNE}} \\
SGCN     & 0.5757 & 0.2626 & 0.5757 & 0.3133 & 0.4636 & 0.2579 & 0.6372 & 0.4493 \\
SIGAT    & 0.5683 & 0.2746 & 0.6512 & 0.3560 & 0.5009 & 0.2585 & 0.6987 & 0.5613 \\
SDGNN    & 0.5785 & 0.3746 & 0.5813 & 0.3660 & 0.4912 & 0.2779 & 0.6765 & 0.5152 \\
SigMaNet & 0.5703 & 0.2918 & 0.5973 & 0.4311 & 0.5472 & 0.3380 & 0.6915 & 0.5033 \\
\midrule
\multicolumn{9}{>{\columncolor{gray!15}}c}{\textbf{DSNE}} \\
SEMBA    & 0.5941 & 0.5958 & 0.6100 & 0.6464 & 0.5447 & 0.5439 & 0.5820 & 0.5801 \\
PolarDSN & \underline{0.7637} & \underline{0.6811} & \underline{0.7182} & \underline{0.7109} & \underline{0.6861} & \underline{0.7177} & \underline{0.7301} & \underline{0.7129} \\
IDP-DSN(ours)  & \textbf{0.8921} & \textbf{0.8406} & \textbf{0.8398} & \textbf{0.8817} & \textbf{0.8928} & \textbf{0.9003} & \textbf{0.8665} & \textbf{0.9187} \\
\bottomrule
\end{tabular}
\end{adjustbox}
\end{table}

\subsubsection{Baselines}
We compare IDP-DSN with 13 competitive baselines in three categories:
\begin{enumerate}[label=(\roman*)]
  \item \textbf{Signed network embedding (SNE)} methods, which learn signed representations on static snapshots
  (e.g., SGCN~\cite{derr2018signed}, SiGAT~\cite{huang2019signed}, SDGNN~\cite{huang2021sdgnn}, SigMaNet~\cite{fiorini2023sigmanet});
  \item \textbf{Dynamic network embedding (DNE)} methods, which are designed for (mostly unsigned) temporal graphs and are adapted to signed graphs by treating edge polarity/weight as edge attributes
  (e.g., TGAT~\cite{xu2020inductive}, TGN~\cite{rossi2020temporal}, CAW~\cite{wang2021inductive}, PINT~\cite{souza2022provably}, GraphMixer~\cite{cong2023we}, DyFormer~\cite{cong2023dyformer}, TIGER~\cite{zhang2023tiger});
  \item \textbf{Dynamic signed network embedding (DSNE)} methods,
  which explicitly model both the temporal evolution of interactions and the opposing semantics of edge polarities in signed networks
  (e.g., SEMBA~\cite{sharma2023representation}, PolarDSN~\cite{kim2024polardsn}).
\end{enumerate}

\textbf{Adapting SNE/DNE to dynamic edge prediction.}
We train SNE baselines by ordering the learning sequence according to the chronological order of edge occurrences, and
adapt DNE baselines to signed graphs by treating edge polarity as an edge attribute, following PolarDSN~\cite{kim2024polardsn}.

\subsubsection{Implementation Details}
We implement IDP-DSN in PyTorch. For each dataset, we select hyperparameters via grid search on the validation set, using
Validation Macro-F1 as the selection criterion; all reported test results use the best configuration selected for that dataset. We set the embedding and memory dimension to $d=64$ for all datasets. For time modeling, intervals $\Delta t$ are transformed by $\log(1+\Delta t)$ and encoded into a 32-dimensional vector using a 2-layer MLP-based time encoder.
The grid search is conducted per dataset over context and optimization settings, including the number of temporal neighbors, the number and length of random walks, the temporal attention depth and number of heads, and the orthogonality weight $\lambda$ for static--dynamic disentanglement. We restrict the search space to configurations that fit the GPU memory budget for each dataset. We train with AdamW (learning rate $10^{-3}$, weight decay $10^{-3}$) and batch size 64, and apply early stopping with patience 5 based on Validation Macro-F1.
To account for training variance, we train each model five times with different random seeds (0--4) and report mean results. All experiments run on a single machine equipped with one NVIDIA RTX 4090 GPU with 24\,GB memory, an AMD Ryzen 9 7950X3D CPU, and 96\,GB host memory.

\subsection{Overall Performance}
To answer \textbf{RQ1}, Table~\ref{tab:overall} reports results on the \textbf{Hybrid} test set, and
Table~\ref{tab:trans-induc} further separates \textbf{Transductive} and \textbf{Inductive} performance.
We summarize the main observations as follows.

\begin{itemize}
  \item \textbf{Overall ranking.}
  IDP-DSN achieves the best Accuracy (Acc.), Weighted-F1 (W-F1), and Macro-F1 (M-F1) on all four datasets, with consistent gains across metrics.

  \item \textbf{Hybrid-set gains over the strongest DSNE baseline.}
On the Hybrid test set, IDP-DSN improves Macro-F1 over PolarDSN by \textbf{+22.4\%} to \textbf{+27.3\%}.
The largest gain is on Epinions at \textbf{+27.3\%} with an absolute increase of 0.1952, while the smallest gain is on BitcoinAlpha at \textbf{+22.4\%} with an absolute increase of 0.1553.
BitcoinOTC and Wiki-RfA show gains of \textbf{+22.9\%} and \textbf{+25.5\%}, corresponding to absolute improvements of 0.1628 and 0.1827, respectively.

  \item \textbf{What we learn from baseline families.}
  SNE methods perform worst because they ignore temporal order and recency.
  DNE methods improve by modeling event streams, but encoding polarity as an edge attribute can mix positive and negative evidence, often hurting the minority \textsc{Neg} class and reducing Macro-F1.
  DSNE baselines are stronger by explicitly modeling signed temporal processes; IDP-DSN further improves them via polarity-specific histories and time-respecting higher-order context, which is beneficial under sparse observations and continual node arrivals.

  \item \textbf{Scalability under the same GPU budget.}
  Under our hardware budget and official configurations, CAW~\cite{wang2021inductive} and PINT~\cite{souza2022provably} exceed GPU memory on Epinions, so we mark those entries as unavailable.
  IDP-DSN fits the same budget and still achieves strong results on Epinions.

  \item \textbf{Inductive setting benefits the most.}
  Inductive Macro-F1 gains over PolarDSN range from \textbf{+23.4\%} to \textbf{+28.9\%}.
  Epinions sees the largest gain at \textbf{+28.9\%} with an absolute increase of 0.2058, while BitcoinOTC is the smallest at \textbf{+24.0\%} with an absolute increase of 0.1708.
  BitcoinAlpha and Wiki-RfA reach \textbf{+23.4\%} and \textbf{+25.4\%}, corresponding to absolute gains of 0.1595 and 0.1826.
  This supports that polarity-aware decoupling and time-respecting structural context improve representations for unseen nodes with short histories while preserving strong transductive performance.
\end{itemize}

\subsection{Ablation Study}
\begin{figure}
  \centering
  \includegraphics[width=\linewidth]{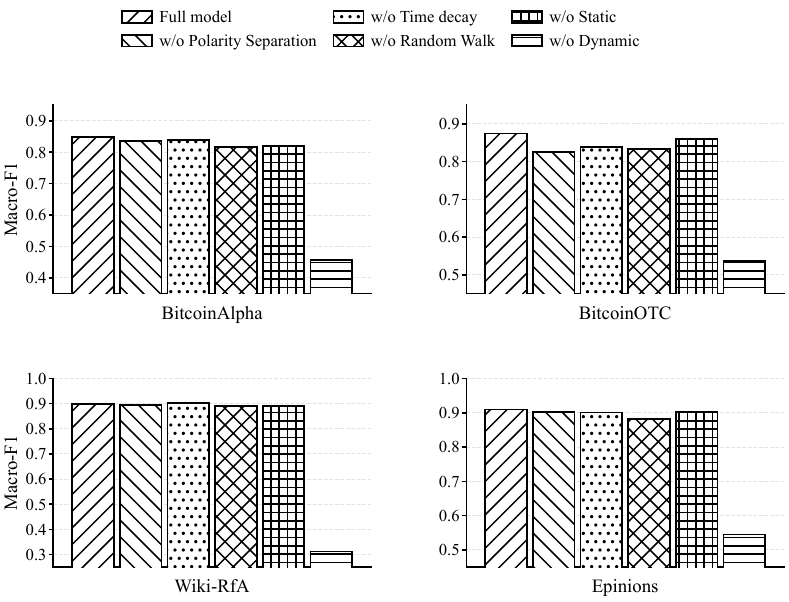}
  \caption{Ablation results (Macro-F1) on the Hybrid test set, which is the union of transductive and inductive test edges.}
  \label{fig:ablation_bar}
\end{figure}

To answer \textbf{RQ2}, we conduct controlled ablations on IDP-DSN.
  We compare the full model with five variants, each removing exactly one component while keeping the backbone, embedding
  dimensions, training protocol, data splits, and evaluation negatives unchanged.
  Specifically, w/o Polarity Separation replaces Dual-Polarity Memory with a single shared memory state,w/o Dynamic removes dynamic representations and uses static embeddings only,w/o Static removes static representations while preserving the full dynamic backbone,w/o Walk Context removes Random Walk Aggregation and its cross-attention pooling,
  and w/o Time-decay disables the learnable decay term in Temporal Neighbor Attention.
  Figure~\ref{fig:ablation_bar} reports Macro-F1 on the Hybrid test set, and Table~\ref{tab:ablation} further breaks down results
  on transductive versus inductive test edges.
  We report the relative drop from the full model as
  $\Delta = \frac{F_{\text{full}}-F_{\text{variant}}}{F_{\text{full}}}\times 100$,
  where $F\in[0,1]$ denotes Macro-F1.

On the Hybrid test set, the same trend is already visible in Fig.~\ref{fig:ablation_bar}.
The largest impact comes from removing dynamics: w/o Dynamic reduces Macro-F1 by \textbf{46.04\%} on BitcoinAlpha,\textbf{38.60\%} on BitcoinOTC, \textbf{65.37\%} on Wiki-RfA, and \textbf{40.15\%} on Epinions.
Removing the static component while preserving the full dynamic backbone (w/o Static) also leads to consistent but much smaller degradations, reducing Macro-F1 by \textbf{3.24\%} on BitcoinAlpha, \textbf{1.60\%} on BitcoinOTC, \textbf{0.97\%} on Wiki-RfA,and \textbf{0.68\%} on Epinions.
In contrast, w/o Random Walk and w/o Polarity Separation lead to smaller but consistent degradations, falling within \textbf{0.86\%--4.72\%} and \textbf{0.61\%--5.60\%}, respectively.
The effect of w/o Time decay is generally modest on Hybrid, with \textbf{0.93\%--4.08\%} drops on three datasets, while Wiki-
RfA is close to neutral and shows a slight gain of \textbf{0.52\%} in this Hybrid view.

Table~\ref{tab:ablation} shows that the dominance of dynamics persists when separating transductive and inductive edges.The w/o Dynamic variant causes a large and consistent performance drop across all datasets.
On inductive edges, Macro-F1 degrades by \textbf{41.70\%--66.66\%}, reaching \textbf{47.64\%} on BitcoinAlpha and \textbf{66.66\%} on Wiki-RfA.
The effect remains substantial in the transductive setting, with a degradation of \textbf{19.33\%--41.01\%}.This indicates that static-only representations are insufficient for dynamic signed prediction, where interaction patterns and signed evidence can change over time.
By contrast, removing the static component while preserving the dynamic backbone (w/o Static) leads to much smaller degradations.On transductive edges, Macro-F1 drops by \textbf{1.32\%--8.88\%}, while on inductive edges the decrease is limited to
\textbf{0.36\%--2.63\%}.
This shows that dynamic modeling is the primary driver of performance, while static information provides additional but smaller gains on top of it.

\begin{table}
  \centering
  \footnotesize
  \setlength{\tabcolsep}{2.5pt}
  \renewcommand{\arraystretch}{1.0}
  \caption{Ablation results (Macro-F1) on \textbf{Transductive} and \textbf{Inductive} test edges. We keep all settings identical and only remove the specified component.}
  \label{tab:ablation}
  \resizebox{\columnwidth}{!}{%
  \begin{tabular}{lcccccccc}
    \toprule
    \multirow{2}{*}{Method}
    & \multicolumn{2}{c}{\textbf{BitcoinAlpha}} & \multicolumn{2}{c}{\textbf{BitcoinOTC}} & \multicolumn{2}{c}{\textbf{Wiki-RfA}} & \multicolumn{2}{c}{\textbf{Epinions}} \\
    \cmidrule(lr){2-3}\cmidrule(lr){4-5}\cmidrule(lr){6-7}\cmidrule(lr){8-9}
     & Trans. & Induc. & Trans. & Induc. & Trans. & Induc. & Trans. & Induc. \\
    \midrule
    IDP-DSN (ALL)   & \textbf{0.8921} & \textbf{0.8406} & \textbf{0.8398} & \textbf{0.8817} & \textbf{0.8928} & \textbf{0.9003} & \textbf{0.8665} & \textbf{0.9187} \\
    w/o polarity separation  & 0.8719 & 0.8217 & 0.8304 & 0.8182 & 0.8834 & 0.8952 & 0.8419 & 0.9151 \\
    w/o time decay  & 0.8659 & 0.8305 & 0.8332 & 0.8384 & 0.8804 & 0.8970 & 0.8209 & 0.9075 \\
    w/o random walk & 0.8689 & 0.8170 & 0.8300 & 0.8266 & 0.8797 & 0.8947 & 0.8469 & 0.9088 \\
    w/o dynamic     & 0.5674 & 0.4401 & 0.6327 & 0.5140 & 0.5267 & 0.3002 & 0.6990 & 0.5171 \\
    w/o static      & 0.8129 & 0.8185 & 0.8249 &
    0.8674 & 0.8810 & 0.8971 & 0.8140 & 0.9122 \\
    \bottomrule
  \end{tabular}%
  }
\end{table}

Once dynamics are enabled, removing the static component leads to much smaller degradations than removing dynamics, indicating that dynamic modeling remains the primary driver of performance, while static information provides complementary gains on top of it.
Beyond this static--dynamic comparison, the remaining modules yield smaller but dependable improvements, more pronounced under induction.
Removing Polarity Separation consistently hurts inductive performance, with the largest drops on the Bitcoin datasets:\textbf{3.49\%} on BitcoinOTC and \textbf{2.25\%} on BitcoinAlpha.
This suggests that separating positive and negative evidence into distinct channels mitigates cross-sign interference when historical supervision is limited. Removing Random Walk also causes a reliable inductive loss (\textbf{0.62\%--2.81\%}), showing that higher-order walk-based signals complement local neighbor evidence for sparse or newly observed nodes.
The contribution of Time decay is modest but consistent: disabling the learnable decay term reduces inductive Macro-F1 by \textbf{0.37\%--1.22\%} across all datasets and causes a larger transductive drop of \textbf{5.26\%} on Epinions, where activity is irregular and time gaps are heterogeneous.Together, these results reveal a clear hierarchy: dynamics drive most gains, static information provides additional but smaller gains, and polarity separation, random-walk aggregation, and time decay provide complementary refinements that primarily enhance inductive generalization.

\subsection{Hyperparameter Sensitivity}
\begin{figure}[t]
  \centering
  \includegraphics[width=\linewidth]{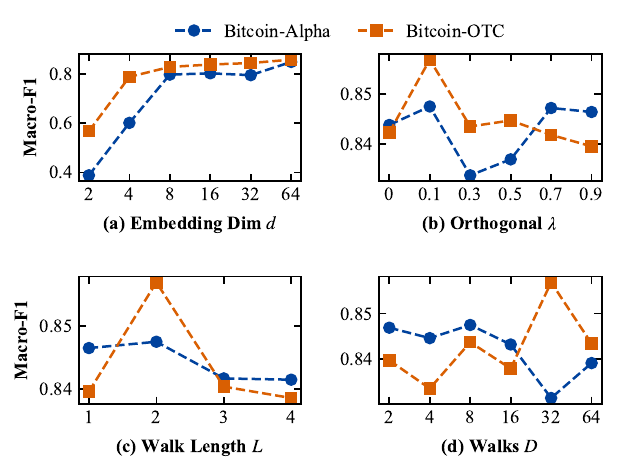}
  \caption{Hyperparameter sensitivity (Macro-F1) of IDP-DSN on dynamic edge prediction (BitcoinAlpha/BitcoinOTC; Hybrid setting).}
  \label{fig:hyperparameter}
\end{figure}

To answer \textbf{RQ3}, Figure~\ref{fig:hyperparameter} studies the sensitivity of Macro-F1 to the embedding dimension $d$, orthogonality weight $\lambda$, walk length $L$, and the number of sampled walks $D$ on BitcoinAlpha and BitcoinOTC under the Hybrid setting. As shown in Fig.~\ref{fig:hyperparameter}(a), performance improves when $d$ is small and then
saturates as the model capacity increases. Fig.~\ref{fig:hyperparameter}(b) shows that IDP-DSN is relatively robust to $\lambda$ across a broad range. Notably, $\lambda=0$ removes the orthogonality regularizer and serves as a no-disentanglement variant. In both datasets, using a moderate $\lambda$ consistently outperforms $\lambda=0$, indicating
that static--dynamic disentanglement is beneficial. Very large $\lambda$ can over-constrain the representations and slightly reduce performance. Regarding walk length, Fig.~\ref{fig:hyperparameter}(c) suggests that short walks are
preferable: $L\!=\!2$ performs best, and longer walks tend to introduce noisier context in signed event streams. Finally,Fig.~\ref{fig:hyperparameter}(d) varies $D$. We find that very small $D$ under-explores the temporal neighborhood,whereas large $D$ brings diminishing returns and increases computational cost. Overall, a moderate $D$ provides a favorable trade-off between accuracy and efficiency, and we use it in the remaining experiments.

\section{Conclusion and Future Work}
For 3-class signed dynamic edge prediction in dynamic signed networks, we propose \textbf{IDP-DSN}, a polarity-aware model that maintains Dual-Polarity Memory with sign-selective updates to reduce cross-sign interference and track evolving interactions. 
To handle sparse histories, IDP-DSN integrates Random Walk Aggregation as multi-hop structural evidence and equips Temporal Neighbor Attention with a learnable time-decay bias for irregular activity patterns. 
Experiments on four signed temporal networks show IDP-DSN consistently outperforms static signed embedding baselines, continuous-time dynamic models adapted to signed settings, and prior dynamic signed methods in both transductive and inductive evaluation. 
Ablation studies confirm the contribution of polarity separation, dynamic representations, random-walk evidence, and time decay, with strongest gains in inductive and sparse-history regimes.

Future work will address current trade-offs to enhance expressiveness and robustness.  
We aim to replace scalar interaction attribute processing with richer feature encodings and non-linear projections for better modeling of complex weighted effects, and refine random-walk sampling by replacing the fixed time-decay bias with adaptive mechanisms that effectively balance recency and long-range structural signals to avoid consistently down-weighting influential past events.  
Additionally, we will explore earlier cross-polarity interactions during memory updates: while late fusion improves stability via explicit polarity separation, controlled early interactions may better capture complex signed dynamics, such as relationship-driven sign reversals.  
The model will also be scaled to larger real-world datasets and extended to heterogeneous signed interactions.

\section{Acknowledgements}
This work is supported by the Natural Science Foundation of China (No. 62402398, No. 62403062, No. 72374173), the University Innovation Research Group of Chongqing (No. CXQT21005), the Technological Innovation and Application Development Project of Chongqing (No. CSTB2025TIAD-KPX0027), the Fundamental Research Funds for the Central Universities (No. SWU-KR24025, No. SWU-XDJH202303) and  the Doctoral Research Start-up Foundation of Guizhou Normal University (No. GZNUD[2026]04).
The experiments are supported by the High Performance Computing clusters  and the  Large-Scale Instrument Sharing Platform (H20 GPU Server) at Southwest University

\bibliographystyle{ACM-Reference-Format} 
\bibliography{refs}

\end{document}